\documentclass[12pt]{article}
\usepackage{graphicx,hyperref,amssymb,amsfonts,amsmath}

\newcommand{\be}{\begin{equation}}
\newcommand{\ee}{\end{equation}}
\newcommand{\bea}{\begin{eqnarray}}
\newcommand{\eea}{\end{eqnarray}}
\newcommand{\beas}{\begin{eqnarray*}}
\newcommand{\eeas}{\end{eqnarray*}}

\begin{document}

\begin{titlepage}

\begin{center}

{\Large Dressing bulk fields in AdS${}_3$}

\vspace{10mm}

\renewcommand\thefootnote{\mbox{$\fnsymbol{footnote}$}}

Daniel Kabat${}^{1}$\footnote{daniel.kabat@lehman.cuny.edu},
Gilad Lifschytz${}^{2}$\footnote{giladl@research.haifa.ac.il}

\vspace{6mm}

${}^1${\small \sl Department of Physics and Astronomy} \\
{\small \sl Lehman College, City University of New York, Bronx NY 10468, USA}

\vspace{4mm}

${}^2${\small \sl Department of Mathematics and} \\
{\small \sl Haifa Research Center for Theoretical Physics and Astrophysics} \\
{\small \sl University of Haifa, Haifa 31905, Israel}

\end{center}

\vspace{15mm}

\noindent
We study a set of CFT operators suitable for reconstructing a charged bulk scalar field $\phi$ in AdS${}_3$ (dual to an operator ${\cal O}$ of dimension $\Delta$
in the CFT) in the presence of a conserved spin-$n$ current in the CFT.  One has to sum a tower of smeared non-primary scalars $\partial_{+}^{m} J^{(m)}$, where
$J^{(m)}$ are primaries with twist $\Delta$ and spin $m$ built from ${\cal O}$ and the current.  The coefficients of these operators can be fixed by demanding that bulk correlators
are well-defined: with a simple ansatz this requirement allows us to calculate bulk correlators directly from the CFT.  They are built from specific polynomials of the kinematic
invariants up to a freedom to make field redefinitions.  To order $1/N$ this procedure captures the dressing of the bulk scalar field by a radial generalized Wilson line.

\end{titlepage}

\setcounter{footnote}{0}
\renewcommand\thefootnote{\mbox{\arabic{footnote}}}

\hrule
\tableofcontents
\bigskip
\hrule

\addtolength{\parskip}{8pt}

\section{Introduction}
According to our current understanding of AdS/CFT \cite{Maldacena:1997re}, perturbative reconstruction of bulk field operators from the point of view of the CFT proceeds in
two stages.  First one obtains the leading-order expression in $1/N$, which we label $\phi^{(0)}$ \cite{Hamilton:2006az}.  It behaves as a free field in AdS.  It can be found by solving the
intersecting modular Hamiltonian equations \cite{Kabat:2017mun} (see also \cite{Jafferis:2015del, Faulkner:2017vdd}) and can sometimes be deduced from symmetry considerations \cite{Aizawa:2014yqa,Nakayama:2015mva}. Then one corrects this expression at subleading orders in $1/N$ by adding multi-trace operators and demanding that the corrected
bulk operator have well-defined correlators \cite{Kabat:2018pbj}.  This approach is reviewed in section \ref{section:iepsilon}.

In this note we are interested in the structure of the CFT operators that are needed to represent bulk scalar fields in the presence of massless spin $n \geq 1$ bulk fields,
including bulk gauge fields and gravity. From previous studies for gauge fields and gravity we know some of the properties of the operators involved, especially when inserted in correlation functions with the boundary field strength $F_{\mu \nu}$ or the boundary Weyl tensor \cite{Kabat:2013wga}. However a full classification of the required CFT operators is still lacking, and the way that the expected dressing singularities arise in correlation functions with a boundary current or energy-momentum tensor remains unclear. From the bulk point of view these dressing
singularities indicate where the gauge or gravitational Wilson line that dresses the bulk field operator ends on the boundary.  For a recent discussion of dressing from the bulk point of view see \cite{Donnelly:2015hta, Donnelly:2016rvo}.

Here we study the situation in AdS${}_3$/CFT${}_2$ where  the work of \cite{Anand:2017dav, Chen:2017dnl, Chen:2019hdv} for the cases $n =1,2$ uncovered a beautiful  and  simple structure to the CFT operators involved.  However this structure was phrased in terms of properties and variables that are special to the Virasoro symmetry of 2-D CFT. We therefore reconsider this case using techniques that could be extended to higher dimensions, and we allow for currents of arbitrary spin $n$.  Our goal is to better understand the structure of the operators involved and the origin of the dressing singularities that are present in correlation functions.

We want to study what types of CFT operators are needed to represent a bulk field as a well-defined CFT operator -- meaning an operator with well-defined correlation
functions -- when inserted in correlators with CFT operators that include conserved higher-spin primaries.  To this end consider a 3-point function
\begin{equation}
\langle \phi(Z,x_1) \overline{\cal O}(x_{2}) {\cal O}_{(h,\bar{h})}(x_3) \rangle
\label{3point-l}
\end{equation}
Here $\phi = \phi^{(0)} + \frac{1}{N} \phi^{(1)}$, where $\phi^{(0)}$ is the lowest-order field written as a smearing of a primary operator ${\cal O}$ and the correction $\phi^{(1)}$ is an
operator that remains to be determined.  As needed $\overline{\cal O}$ denotes the complex conjugate of ${\cal O}$, and ${\cal O}_{(h,\bar{h})}$ is some other primary operator with $({\rm right},\, {\rm left})$ conformal dimensions $(h,\bar{h})$.

The reason the correction $\phi^{(1)}$ is necessary is that $\phi^{(0)}$ by itself does not have a well-defined 3-point function \cite{Kabat:2018pbj}.  As long as ${\cal O}_{(h,\bar{h})}$
does not obey a conservation equation we know how to fix this problem \cite{Kabat:2011rz, Kabat:2012av, Kabat:2015swa}.  In the absence of a conservation law we can build a tower of primary scalars of the form
$(\partial)^{i}{\cal O}(\partial)^{j}{\cal O}_{(h,\bar{h})}$, and we can assemble $\phi^{(1)}$ as a sum of these smeared primaries.  The coefficients can be determined
by requiring that the output is a CFT operator $\phi$ whose
3-point function (\ref{3point-l}) is well-defined in a manner described in section \ref{section:iepsilon}.  It turns out this procedure generates a 3-point function that is non-singular at bulk space-like separation.  In this sense bulk microcausality is a consequence of the reconstruction procedure.

If ${\cal O}_{(h,\bar{h})}$ obeys a conservation equation one cannot in general build primary\footnote{Here and in the rest of the paper primary means (even in $d=2$)
primary with respect to the global $SO(d,2)$ conformal group.} scalars from expressions of the form $(\partial)^{i}{\cal O}(\partial)^{j}{\cal O}_{(h,\bar{h})}$. This has important consequences.
It will turn out that we can still build a well-defined CFT operator, but its 3-point function will be singular at bulk spacelike separation. This singularity is a manifestation of the bulk dressing that
must end somewhere on the boundary, and in this sense bulk dressing is a consequence of the reconstruction procedure.  The goal of this paper is to better understand the origin of these dressing singularities from the point of view of the CFT.

To further illustrate the role of the operators we construct, consider a correlator with some number of insertions of a conserved primary spin-one current $j_{-}=j_{(1,0)}$.
\begin{equation}
\langle\phi(Z,x_1^+,x_1^-) {\overline{\cal O}}(x^{+}_2,x_{2}^{-}) j_{-}(x^{-}_{3})\cdots j_{-}(x^{-}_{k})\rangle
\label{3pointj-n}
\end{equation}
We can use the OPE between the currents to convert this into an infinite sum of 3-point functions of the form (\ref{3point-l}), as was done for scalars in \cite{Kabat:2016zzr}. In the process new operators will appear, for instance the multi-trace operator $\big(j_{-})^{k-2}$.  To leading order in $1/N$ this
operator has dimension $k-2$ and spin $k-2$ and is therefore a conserved higher-spin current. At subleading orders in $1/N$ this operator will acquire an anomalous dimension
and will no longer be conserved.  But working order-by-order in the $1/N$ expansion it seems the bulk scalar field in (\ref{3pointj-n}) will have to be corrected by adding non-primary scalars built from $\big(j_-\big)^{k-2}$.  So it seems the corrections we study in this paper will be relevant for understanding multi-point correlators, even if the bulk theory only has fields with spin one.

\section{The $i\epsilon$ issue \label{section:iepsilon}}
As discussed above, our guiding principle is to construct a well-defined CFT operator whose leading term in the $1/N$ expansion obeys the intersecting modular Hamiltonian
equations.  For scalar interactions in the bulk this was studied in \cite{Kabat:2018pbj}.  Here we'd like to extend the analysis to a bulk theory with gauge symmetry.  To do this we work
in the simplest setting of a massless complex scalar field in AdS${}_3$ coupled to a Chern-Simons gauge field.  Thus in the CFT we consider a complex scalar primary ${\cal O}$ with
dimension $\Delta = 2$ that's charged under a right-moving chiral current $j_-(x^-)$.
The 2-point correlators are
\begin{eqnarray}
\langle {\cal O}(x_1) {\overline{\cal O}}(x_2) \rangle = {1 \over \left(x_{12}^+ x_{12}^-\right)^2} \\
\langle j_-(x_1^-) j_-(x_2^-) \rangle = - {1 \over \left(x_{12}^-\right)^2}
\end{eqnarray}
and the 3-point correlator is
\be
\label{CFT3pt}
\langle {\cal O}(x_1) \overline{\cal O}(x_2) j_-(x_3) \rangle = - {\lambda \over 2} {1 \over (x_{12}^2)^2} \left(- {1 \over x_{13}^-} + {1 \over x_{23}^-}\right)
\ee
Here $x^\pm = t \pm x$ and $x_{ij} = x_i - x_j$.

\begin{figure}
\begin{center}
\includegraphics{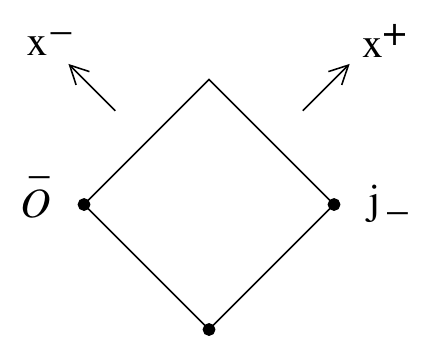}
\end{center}
\caption{$\overline{\cal O}$ and $j_-$ define a causal diamond on the boundary.  As expected, the 3-point function (\ref{phi0Oj}) has a light-cone singularity when the bulk point is null separated from $\overline{\cal O}$.  But it also has an ill-defined singularity when the bulk point is null separated from the bottom corner of the causal diamond.\label{fig:diamond}}
\end{figure}

The solution of the intersecting modular Hamiltonian equations is \cite{Kabat:2017mun}
\begin{equation}
\label{masslessphi0}
\phi^{(0)}(t,x,Z) = {1 \over 2\pi} \hspace{-3mm} \int\limits_{\hspace{5mm} (t')^2 + (y')^2 < Z^2} \hspace{-5mm} dt'dy' \, {\cal O}(t + t',x + iy').
\end{equation}
Using this to move the first operator in (\ref{CFT3pt}) into the bulk, the resulting 3-point correlator is
\be
\langle \phi^{(0)}(x_1,Z) \overline{\cal O}(x_2) j_-(x_3) \rangle = - {\lambda \over 4 \pi i} \int_0^Z rdr \oint_{\vert \alpha \vert = 1} \alpha d\alpha
{1 \over (x_{12}^+ + r \alpha)^2} {1 \over (\alpha x_{12}^- + r)^2} \left(-{\alpha \over \alpha x_{13}^- + r} + {1 \over x_{23}^-}\right)
\ee
where 
\be
t' = r \cos \theta, \qquad y' = r \sin \theta,  \qquad \alpha = e^{i \theta}.
\ee
Performing the contour integral and then the $r$ integral we find
\be
\label{phi0Oj}
\langle \phi^{(0)}(x_1,Z) \overline{\cal O}(x_2) j_-(x_3) \rangle = - {\lambda Z^2 \over 4} {1 \over (-x_{12}^+x_{12}^-+Z^2)(-x_{12}^+x_{13}^- + Z^2) x_{23}^-}
\ee
This correlator only has poles.
\begin{enumerate}
\item
There's a pole at $x_{12}^+x_{12}^- = Z^2$, when the bulk operator is at null bulk separation from $\overline{\cal O}$.
\item
There's a pole at $x_{12}^+x_{13}^- = Z^2$.  If $\overline{\cal O}$ and $j_-$ are spacelike separated on the boundary, with $\overline{\cal O}$ to the left, then this
singularity is at null bulk separation from the bottom corner of the boundary causal diamond illustrated in Fig.\ \ref{fig:diamond}.
\item
There's a pole at $x_{23}^- = 0$, when $\overline{\cal O}$ and $j_-$ are null separated along the $x^+$ direction on the boundary.
\end{enumerate}

To define the CFT correlator (\ref{CFT3pt}) we use a Wightman $i \epsilon$ prescription, namely
\be
t_i \rightarrow t_i - i \epsilon_i \qquad\qquad \hbox{\rm }
\ee
with operators ordered from left to right in the correlator by decreasing values of $\epsilon_i$.  The question is whether this prescription in the CFT
is sufficient to make the bulk correlator (\ref{phi0Oj}) well-defined.
We find that the first and third poles acquire sensible $i \epsilon$ prescriptions, but the $i \epsilon$ prescription
for the second pole is ambiguous if $\epsilon_{12}$ and $\epsilon_{13}$ have different signs -- that is, if the bulk operator is in the middle of the correlator.
In that case it's ambiguous whether the pole is displaced into the upper or lower half plane (for further discussion of this point see \cite{Kabat:2018pbj}). This means that
$\phi^{(0)}$ -- although it behaves sensibly inside a 2-point function -- is not a well-defined operator once one considers its behavior inside 3-point correlators.

In the rest of this paper we will see how to cure this problem.

\section{CFT computation of the corrected bulk operator \label{section:main}}
As explained in the introduction our procedure for bulk reconstruction is to build a CFT operator whose leading term in the $1/N$ expansion obeys the intersecting modular Hamiltonian
equations, with $1/N$ corrections that are chosen so the full bulk operator gives unambiguous results when inserted in correlation functions.  Here we use this procedure to build a bulk
scalar field in AdS${}_3$.

We start with the CFT three-point function of a pair of primary scalars with a conserved spin-$n$ primary current.
\begin{equation}
\label{OObarj}
\langle{\cal O}(x_{1}^{+},x_{1}^{-})\overline{\cal O}(x_{2}^{+},x_{2}^{-})j_{(n,0)}(x_{3}^{-})\rangle=\frac{\gamma_{n}}{(x_{12}^{-})^{\Delta-n}(x_{12}^{+})^{\Delta}(x_{13}^{-})^{n}(x_{23}^{-})^{n}}
\end{equation}
Here ${\cal O}$ is a primary scalar with conformal dimension $\Delta$ and $\overline{O}$ denotes its complex conjugate as needed.\footnote{${\cal O}$ needs to be complex when
$n$ is odd.} $j_{(n,0)}(x^-)$ is a conformal primary of dimension $(n,0)$ and $\gamma_n$ is an OPE coefficient.

The zeroth-order expression for the bulk field is \cite{Hamilton:2006az}
 \begin{equation}
\phi^{(0)}(Z,t,x) = \hspace{-3mm} \int\limits_{\hspace{5mm} (t')^2 + (y')^2 < Z^2} \hspace{-5mm} dt'dy' K_{\Delta} \, {\cal O}(t + t', x + i y')
\label{smearing}
\end{equation}
where
\begin{equation}
K_\Delta = {1 \over 2 \pi} \left({Z^2 - (t')^2 - (y')^2 \over Z}\right)^{\Delta-2}
\end{equation}
This provides a solution to the intersecting modular Hamiltonian equations for the CFT vacuum \cite{Kabat:2017mun}, normalized so that
\begin{equation}
\phi^{(0)}(Z,x) \simeq {Z^\Delta \over 2\Delta - 2} {\cal O}(x) \qquad \hbox{\rm as $Z \rightarrow 0$}
\end{equation}
Using this representation to move the first operator in (\ref{OObarj}) into the bulk we find (see appendix \ref{appendix:geq})
\begin{equation}
\langle\phi^{(0)}(Z, x_{1}^{+},x_{1}^{-})\overline{\cal O}(x_{2}^{+},x_{2}^{-})j_{(n,0)}(x_{3}^{-})\rangle=\frac{\tilde{\gamma}_{n}}{(x_{23}^{-})^{n}Z^{\Delta}(1-\chi_1)^{\Delta-n}(1-\chi_2)^n}
\label{phi0a}
\end{equation}
where $\tilde{\gamma}_n = \gamma_n/(2\Delta - 2)$ and where we've introduced the two kinematic invariants\footnote{Invariant under the bulk lift of Poincar\'e and scale transformations in the CFT, but not under the bulk lift of special conformal transformations.}
\begin{equation}
x^{\pm}=t \pm x, \ \ x_{ij}=x_i -x_j,\ \  \chi_{1}=\frac{x_{12}^{+} x_{12}^{-}}{Z^2}, \ \ \chi_{2}=\frac{x_{12}^{+} x_{13}^{-}}{Z^2}
\end{equation}
The problem with the result (\ref{phi0a}) is the pole at $\chi_{2}=1$, since this pole suffers from the ambiguous $i \epsilon$ prescription described
in section \ref{section:iepsilon} for the case $n = 1$.  The other singularities in (\ref{phi0}) do not suffer from any ambiguity and hence are not an obstacle to thinking of $\phi^{(0)}$ as a
well-defined CFT operator.  It will be useful to introduce the combination
\begin{equation}
Y=\frac{\chi_{2}-\chi_{1}}{\chi_{2}(1-\chi_1)} = \frac{- Z^2 x_{23}^{-}}{x^{-}_{13}(x^{+}_{12}x^{-}_{12} - Z^2)}
\end{equation}
and re-write the correlator as
\begin{equation}
\langle\phi^{(0)}(Z, x_{1}^{+},x_{1}^{-})\overline{\cal O}(x_{2}^{+},x_{2}^{-})j_{(n,0)}(x_{3}^{-})\rangle=\frac{(-1)^{n} \, \tilde{\gamma}_{n}}{(x_{23}^{-})^{n}Z^{\Delta}(1-\chi_1)^{\Delta}}\left(\frac{\chi_1}{\chi_2}\right)^{n}\frac{1}{(Y-1)^{n}}
\label{phi0}
\end{equation}
In this form the problematic pole is at $Y = 1$; despite appearances there is no pole at $\chi_2 = 0$.

To cure this problem we correct our CFT definition of the bulk field to
\be
\phi = \phi^{(0)} + {1 \over N} \phi^{(1)}
\ee
We expect that $\phi^{(1)}$ should be a double-trace operator built from ${\cal O}$ and $j_{(n,0)}$, chosen to make the corrected correlation function $\langle \phi \, \overline{\cal O} \, j_{(n,0)} \rangle$ well-defined.  There is a huge
freedom in the choice of $\phi^{(1)}$, corresponding to the freedom to choose a gauge and a dressing for the bulk field (as well as to make bulk field redefinitions).  Although it would be
interesting to proceed in full generality, by searching for the most general correction $\phi^{(1)}$ that makes the bulk field well-defined, as a practical matter it is convenient to restrict the
space of CFT operators we consider.\footnote{General dressings have been considered in AdS${}_3$ in \cite{Chen:2019hdv}.}  We will attempt to build $\phi^{(1)}$ out of CFT operators ${\cal A}$
that are Lorentz scalars and have definite scaling dimensions.  However we will {\em not} require that ${\cal A}$ transform as a conformal primary, for the simple reason that it's not possible to build a
scalar primary out of ${\cal O}$ and a conserved current.  Instead we will require that under a special conformal transformation with parameter $b^\mu$
\begin{equation}
\label{perpcond}
\delta_{b} {\cal A} \sim (\hbox{\rm primary scalar result}) + \partial \cdots \partial (b^{\mu} j_{\mu \cdots}) \partial \cdots \partial {\cal O}
\end{equation}
That is, we require that $b^\mu$ be contracted with an index on the current and we forbid contractions between $b^\mu$ and a derivative operator.
In a 2-D CFT with only $j_{-}$ present this condition means the transformation of ${\cal A}$ under $L_{1}$ is that of a primary scalar but the transformation
under $\overline{L}_{1}$ is not.

One motivation for the condition (\ref{perpcond}) is just that it will allow us to obtain simple and explicit expressions for the corrected bulk operator.  In fact
operators obeying (\ref{perpcond}) have been used before to cancel unwanted singularities in correlation functions involving a bulk scalar field and a boundary field strength
$F_{\mu\nu} = \partial_\mu j_\nu - \partial_\nu j_\mu$ \cite{Kabat:2012av,Kabat:2013wga}.  As further motivation we recall that
for spin-1 currents operators obeying (\ref{perpcond}) in the CFT are expected to appear when fixing holographic gauge $A_Z = 0$ in the bulk \cite{Kabat:2012av,Kabat:2013wga}.  To see this recall that holographic gauge $A_Z = 0$ is preserved by Poincar\'e and scale transformations in the CFT,
but under an infinitesimal special conformal transformation \cite{Kabat:2012hp}
\bea
\label{AZprime}
&& \delta_b A_Z = 2 Z b \cdot A \\
\nonumber
&& \delta_b A_\mu = 2 x_\mu b \cdot A - 2 b_\mu x \cdot A - 2 b \cdot x A_\mu
\eea
Thus we need a compensating gauge transformation with
\be
\label{comp}
\delta A_Z = \partial_Z \lambda = - 2 Z b \cdot A
\ee
to restore holographic gauge, and as a result a charged bulk scalar field acquires an anomalous transformation law \cite{Kabat:2013wga}
\be
\label{trans}
\delta_b \phi = (\hbox{\rm scalar result}) - {i \over N} \lambda \phi\,.
\ee
At leading order in the $1/N$ expansion $A_\mu \sim \partial \cdots \partial j_\mu$ and $\phi^{(0)} \sim \partial \cdots \partial {\cal O}$, which means that (\ref{comp}) involves the combination $b \cdot j$ and (\ref{trans}) has the form
\begin{equation}
\delta_{b} \phi \sim (\hbox{\rm scalar result}) + {1 \over N} \partial \cdots \partial (b^{\mu} j_{\mu}) \partial \cdots \partial {\cal O}
\end{equation}
This bulk result has the same form as (\ref{perpcond}).  So to ${\cal O}(1/N)$ we anticipate that the CFT condition (\ref{perpcond}) corresponds to fixing holographic gauge in the bulk.\footnote{Similar
expressions are available for metric perturbations in Fefferman-Graham gauge \cite{Kabat:2012hp} and were extended to the full Virasoro algebra in \cite{Anand:2017dav}.}  Below
by explicit construction we will see that this is indeed the case.

In searching for CFT operators obeying (\ref{perpcond}) a convenient starting point is to note that from ${\cal O}$, $j_{(n,0)}$ and $l$ $\partial_-$ derivatives one can build
a unique double-trace spinning conformal primary $J_{(l+\frac{\Delta}{2}+n,\frac{\Delta}{2})}(x^+,x^-)$.  As indicated the $({\rm right},\,{\rm left})$ conformal dimensions of $J$
are $(l+\frac{\Delta}{2}+n,\frac{\Delta}{2})$.  This operator has a scalar descendant $\partial_+^{l+n} J_{(l+\frac{\Delta}{2}+n,\frac{\Delta}{2})}$, and one can check that under a special
conformal transformation this scalar descendant obeys (\ref{perpcond}).  These properties are checked explicitly in appendix \ref{appendix:hsp}.

To show that these operators can be used to construct a well-defined bulk field -- that is, that they can be used to cancel the ambiguous singularities in (\ref{phi0}) -- we define
\begin{equation}
{\cal A}^{\Delta,n}_{l}(Z, x^{+},x^{-}) = \int d^2x' \, K_{\Delta+2l+2n}(Z,x \vert x') \partial_{+}^{l+n} J_{(l+\frac{\Delta}{2}+n,\frac{\Delta}{2})}(x')
\label{AN}
\end{equation}
The relevant 3-point function is calculated in appendix \ref{appendix:geq}.
\begin{eqnarray}
\label{An}
& & \langle{\cal A}^{\Delta,n}_{l}(Z, x_{1}^{+},x_{1}^{-})\overline{\cal O}(x_{2}^{+},x_{2}^{-})j_{(n,0)}(x_{3}^{-})\rangle \\
\nonumber
& & = \frac{\tilde{\alpha}^{(n)}_{l}}{(x_{23}^{-})^{n}Z^{\Delta}(1 - \chi_1)^{\Delta}}\left(\frac{\chi_1}{\chi_2}\right)^{n}\left(\frac{Y}{Y-1}\right)^{n}(-Y)^{l}F(l+\Delta, l+n, 2l+2n+\Delta,Y)
\end{eqnarray}
Then we can set
\begin{equation}
\phi^{(1)}=\sum_{l=0}^{\infty} a_{l} {\cal A}^{\Delta,n}_{l}
\label{phi1}
\end{equation}
and try to find coefficients $a_{l}$ so that the combination $\phi^{(0)}+\frac{1}{N}\phi^{(1)}$ is a well-defined CFT operator.

To make the bulk field well-defined we need to cancel the pole at $Y=1$
in (\ref{phi0}) without introducing any new trouble. Note that (\ref{An}) has a branch cut at $1 < Y < \infty$ which inherits an ambiguous $i\epsilon$ prescription from the CFT.
So what we need to do is find coefficients $a_{l}$ in (\ref{phi1}) such that the branch cuts cancel after summing over $l$ while a surviving pole at $Y = 1$ cancels the problematic pole in (\ref{phi0}).
It is indeed possible to do this, as we show explicitly for the case $n=1$ in appendix \ref{appendix:n=1}. Another approach is to use the bulk equations of motion to compute the corrected bulk operator, and we take this approach in appendix \ref{appendix:bulkeom}. Both methods give the same result. However the first approach becomes more and more algebraically involved for
larger values of $n$, while for the second approach one needs to know the bulk theory.  So we will instead follow a different path.

The approach we take amounts to making a minimal ansatz for the analytic properties of correlators.  Combining (\ref{An}) and (\ref{phi1}) we have
\begin{equation}
\langle \phi^{(1)}(Z, x_{1}^{+},x_{1}^{-})\overline{\cal O}(x_{2}^{+},x_{2}^{-})j_{(n,0)}(x_{3}^{-})\rangle =  \frac{1}{(x_{23}^{-})^{n}Z^{\Delta}(1 - \chi_1)^{\Delta}}\left(\frac{\chi_1}{\chi_2}\right)^{n}\left(\frac{Y}{Y-1}\right)^{n}f_{n}(Y)
\label{3pointphi1}
\end{equation}
where
\begin{equation}
\label{fn}
f_n(Y) = \sum_{l=0}^{\infty} a_{l} \tilde{\alpha}_l^{(n)} (-Y)^{l}F(l+\Delta, l+n, 2l+2n+\Delta,Y)
\end{equation}
Comparing to (\ref{phi0}) we see that it is useful to parametrize $f_{n}(Y)$ as 
\begin{equation}
\label{parametrizef}
f_{n}(Y)=\tilde{\gamma}_{n}\left(\frac{(-1)^{n+1}}{Y^{n}}+\left(\frac{Y-1}{Y}\right)^{n}g_{n}(Y)\right)
\end{equation}
so that the corrected 3-point function is 
\begin{equation}
\langle\phi(Z, x_{1}^{+},x_{1}^{-})\overline{\cal O}(x_{2}^{+},x_{2}^{-})j_{(n,0)}(x_{3}^{-})\rangle =  \frac{\tilde{\gamma}_{n}}{(x_{23}^{-})^{n}Z^{\Delta}(1 - \chi_1)^{\Delta}}\left(\frac{\chi_1}{\chi_2}\right)^{n}g_{n}(Y)
\label{3pointphi}
\end{equation}

Now let's consider analyticity.  The branch cuts in the sum (\ref{fn}) are supposed to cancel, so a minimal assumption is that $f_n(Y)$ is an entire function of $Y$.  As the simplest possibility, we take $f_n(Y)$ to be a polynomial.\footnote{Any polynomial will do,
so there are infinitely many solutions to the requirement that the cuts cancel.}  To close a potential loophole, in appendix \ref{appendix:coef} we show that any polynomial $f_n$ can be obtained by appropriately choosing the coefficients $a_l$ in (\ref{fn}).
This takes care of the cuts.  But still, to avoid an ambiguous pole, the corrected correlator (\ref{3pointphi}) should not be singular at $Y = 1$.  Generically if $f_n$ is a polynomial then from (\ref{parametrizef}) $g_n$ will have a singularity at $Y = 1$.
However these two properties -- that $f_n$ is a polynomial and that $g_n$ has no singularity at $Y = 1$ -- can both be satisfied provided $g_{n}(Y)$ is a special polynomial of at least degree $n-1$.  To analyze the simplest possibility (a polynomial of degree
$n-1$) we set
\begin{equation}
\label{gnsum}
g_{n}(Y)=\sum_{j=0}^{n-1} c_{j} Y^{j}
\end{equation}
The requirement that $f_{n}(Y)$ is a polynomial gives conditions on the coefficients $c_{j}$, namely that $c_0 = 1$ and
\begin{equation}
\sum_{j=0}^{m}(-1)^{j}c_{m-j}{n \choose j}=0 \qquad \hbox{\rm for $1\leq m \leq n-1$}
\label{gpolycon}
\end{equation}
We solve this system in appendix \ref{appendix:gnY} to find
\begin{equation}
\label{clsol}
c_{l}={n+l-1 \choose l}
\end{equation}
This is the simplest possibility that gives a well-defined correlator.  There is of course the freedom to make a non-minimal choice and add higher powers of $Y$ to $g_{n}(Y)$; we will see what this freedom means later.

We illustrate these results with a few examples. For $n=1$ the minimal choice for $g_{n}(Y)$ is $g_{1}(Y)=1$.  This gives the known bulk-boundary 3-point function with a conserved vector \cite{Chen:2019hdv}. For $n=2$ the minimal polynomial is $g_{2}(Y)=1+2Y$. Again this agrees with the known result for a 3-point function with the stress tensor \cite{Anand:2017dav}. For $n=3$ the minimal polynomial is  $g_{3}(Y)=1+3Y+6Y^2$ and for $n=4$ the minimal polynomial is $g_{4}(Y)=1+4Y+10Y^2+20Y^3$. We give a bulk
computation reproducing these results in section \ref{section:bulkcomputation}, where we show that for all $n$ the general result obeys the expected bulk equation of motion.

It's worth stressing that the corrected 3-point function (\ref{3pointphi}) is singular when $x^{-}_{13}=0$, while the uncorrected correlator (\ref{phi0}) did not have such a singularity.\footnote{This can be seen explicitly in (\ref{phi0a}), where the only
singularities are at $\chi_1 = 1$, $\chi_2 = 1$, $x_{23}^- = 0$.}  To see this it's convenient to rewrite the corrected correlator (\ref{3pointphi}) as
\begin{equation}
\label{rewrite3pointphi}
\langle \phi(Z,x_1) \overline{\cal O}(x_2) j_{(n,0)}(x_3) \rangle = {\tilde{\gamma}_n Z^\Delta \over (x_{12}^+ x_{12}^- - Z^2)^\Delta} \left({x_{12}^- \over x_{13}^- x_{23}^-}\right)^n g_n\left({-Z^2 x_{23}^- \over x_{13}^- (x_{12}^+ x_{12}^- - Z^2)}\right)
\end{equation}
Since we found that $g_n(Y)$ must be entire (in fact a polynomial) the only singularities are at $x_{12}^+ x_{12}^- = Z^2$, $x_{13}^- = 0$ and $x_{23}^- = 0$.

The singularity at $x^{-}_{13}=0$ has a physical interpretation.  Since the singularity of the bulk field with the current is at $x_3^- = x_1^-$ and is independent of $x_3^+$, we see that the corrected bulk field $\phi(Z,x_1^+,x_1^-)$ produces a right-moving charge on the boundary at $x^- = x_1^-$.
This is consistent with the interpretation that the corrected bulk field is dressed by a generalized Wilson line that runs radially (purely in the $Z$ direction) to the boundary.  The CFT operators (\ref{AN}) we used to construct $\phi^{(1)}$ made
this behavior inevitable since they led to polynomial $g_n$.  By allowing more general operators (i.e.\ not restricting to operators satisfying (\ref{perpcond})) we could have obtained more general dressings for the bulk field \cite{Chen:2019hdv}.

\subsection{Non-minimal choices for $g_{n}(Y)$}
Finally let's examine the freedom to make non-minimal choices for $g_{n}(Y)$ by adding higher powers of $Y$ to the minimal polynomial.  For simplicity we look at the case $n=1$.  Instead of the minimal choice
$g_{1}(Y) =1$, suppose we extend the sum in (\ref{gnsum}) and take $g_{1}(Y) = 1 + c_1 Y$.  From (\ref{3pointphi}) the change in the 3-point function due to the term linear in $Y$ is
\begin{equation}
\frac{c_1 \tilde{\gamma}_1}{x_{23}^{-}Z^{\Delta}(1 - \chi_1)^{\Delta}}\frac{\chi_1}{\chi_2}Y = {c_1 \tilde{\gamma}_1 x_{12}^- \over Z^\Delta (1-\chi_1)^{\Delta + 1} (x_{13}^-)^2}
\label{n=1correc}
\end{equation}
One might suspect that this freedom is related to a field redefinition, as was discussed for scalar interactions in \cite{Kabat:2015swa}. Indeed if we re-define the bulk field as
\begin{equation}
\phi \rightarrow \phi +\frac {c_1\tilde{\gamma}_1}{N\Delta} A^{(0)}_{\mu} \nabla^{\mu} \phi^{(0)}
\end{equation}
then the second term inside a 3-point function with $\overline{\cal O}$ and $j_{-}$ will reproduce (\ref{n=1correc}).
 
\section{Bulk computation of the 3-point function \label{section:bulkcomputation}}
Here we present some bulk computations that reproduce the CFT result (\ref{3pointphi}) for the 3-point function.  We do this in the context of a bulk higher-spin gauge field ${\bf{A}}^{(n)}$, dual to a totally symmetric, traceless, conserved rank-$n$ tensor primary
current $j^{(n)}$ in the CFT.  In particular $j^{(n)}$ obeys
\begin{equation}
\partial^{\nu}j^{(n)}_{\nu \mu_2 \cdots \mu_{n}} = 0 \qquad\quad  j^{(n) \nu}_{\nu \mu_2 \cdots \mu_{n}}=0.
\end{equation}
For a 2-D CFT this means there are only two non-zero components $j_{(n,0)}(x^{-})$ and $j_{(0,n)}(x^{+})$.
In holographic gauge we set ${\bf{A}}^{(n)}_{Z\cdots}=0$ and the higher-spin field can be represented as a smearing of the current \cite{Sarkar:2014dma}.\footnote{We follow the normalizations used in \cite{Sarkar:2014dma}.}
\begin{equation}
{\bf A}^{(n)}_{\mu_1 \cdots \mu_n}(Z,t,{\bf x}) = \frac{\Gamma(n+\frac{d}{2}-1)}{\pi^{\frac{d}{2}}\Gamma(n-1)Z^n} \hspace{-3mm} \int\limits_{\hspace{5mm} (t')^2 + \vert {\bf y}'\vert^2 < Z^2} \hspace{-5mm} dt'd^{d-1}y' \left({Z^2 - (t')^2 - \vert {\bf y}' \vert^2 \over Z}\right)^{n-2} j^{(n)}_{\mu_1\cdots \mu_{n}}(t+t',{\bf x} + i {\bf y'})
\end{equation}
The bulk field inherits the properties
\begin{equation}
\partial^{\nu}{\bf A}^{(n)}_{\nu \mu_2 \cdots \mu_{n}}=0 \qquad\quad  {\bf A}^{(n) \nu}_{\nu \mu_2 \cdots \mu_{n}} = 0
\end{equation}
from the current.
In AdS${}_3$ this means that in holographic gauge the only non-zero components are ${\bf A}^{(n)}_{-\cdots -}(Z,x^{-})$ and ${\bf A}^{(n)}_{+\cdots +}(Z,x^{+})$.  In AdS${}_3$ by evaluating the smearing integral in polar coordinates
we have
\begin{equation}
{\bf A}^{(n)}_{-\cdots -}(Z, x^{+},x^{-})=j_{(n,0)}(x^{-}), \qquad {\bf A}^{(n)}_{+\cdots +}(Z, x^{+}, x^{-})=j_{(0,n)}(x^{+}).
\end{equation}
The coupling of ${\bf A}^{(n)}$ to a bulk scalar (a complex scalar if $n$ is odd) is through a conserved current made out of the scalar field.  For AdS${}_{3}$ in holographic gauge the bulk interaction can be
put in the form (after integration by parts)
\begin{equation}
\sim Z^{2n} {\bf A}^{(n)}_{-\cdots -} \, \phi \partial_{+}^{n} \phi.
\label{bulkvertex}
\end{equation}
That is, the equation of motion for the bulk scalar has the form
\begin{equation}
\label{bulkHSeom}
\big(\nabla^{2}-\Delta(\Delta-2)\big) \phi \sim \frac{1}{N}Z^{2n} {\bf A}^{(n)}_{-\cdots -} \partial_{+}^{n} \phi
\end{equation}

With this bulk description in hand we can test the CFT result (\ref{3pointphi}) by showing that the equation of motion for the bulk field (\ref{bulkHSeom}) holds inside a 3-point function.
That is, we should have
\begin{equation}
\big(\nabla^{2}_{Z,x_1}-\Delta(\Delta-2)\big)\langle\phi(Z,x_{1}) \overline{\cal O}(x_2) j_{(n,0)}(x^{-}_3)\rangle\sim \frac{1}{N}Z^{2n}\langle {\bf A}^{(n)}_{-\cdots -} \partial_{+}^{n} \phi(Z,x_{1})\overline{\cal O}(x_2) j_{(n,0)}(x^{-}_3)\rangle
\label{eomsec}
\end{equation}
Using (\ref{3pointphi}) to evaluate the left-hand side, we show that this equation of motion is indeed satisfied in appendix \ref{appendix:eom}.

As an alternate way of verifying (\ref{3pointphi}), let us evaluate a tree-level Witten diagram for 
\begin{equation}
\langle \phi(Z,x_1)\overline{\cal O}(x_2) j_{(n,0)}(x^{-}_3)\rangle
\end{equation}
We denote the bulk-bulk scalar Feynman propagator by $G_{\phi}$, the bulk-boundary scalar propagator by $K_{\phi}$ and the bulk-boundary higher-spin propagator by $K_{A}$.  With ${\bf X}$ representing
a bulk point the latter is given by
\begin{equation}
K_{A}({\bf X},x) \sim \frac{1}{({\bf X}^{-}-x^{-})^{2n}}
\end{equation}
The bulk-boundary 3-point function is then given by
\begin{equation}
\langle\phi({\bf X}_1) \overline{\cal O}(x_2) j_{-\cdots -}^{(n)}(x^{-}_3)\rangle \sim  \int d^3 {\bf X}' \sqrt{g} \, (Z')^{2n} G({\bf X}_1,{\bf X}')(\partial_{+}^{n}K_{\phi}({\bf X}',x_2))K_{A}({\bf X}',x_3).
\end{equation}
We follow \cite{Anand:2017dav,Chen:2019hdv} and evaluate the integral in the approximation that the conformal dimension of the scalar operator is large.  This gives
\begin{equation}
\label{bulk3pt}
\langle\phi(Z,0,0) \overline{\cal O}(x_2) j_{(n,0)}(x^{-}_3)\rangle \sim  \langle\phi(X_1) \overline{\cal O}(x_2)\rangle\int_{0}^{x^{-}_{2}}dx^{-}\frac{1}{(x^{-}-x^{-}_{3})^2}\left( \frac{(x^{-}_{2}-x^{-})(Z^2-x^{-}x_{2}^{+})}{(x^{-}_{3}-x^{-})^2(Z^2-x_{2}^{-}x_{2}^{+})}\right)^{n-1}
\end{equation}
This will agree with the CFT result (\ref{3pointphi}) provided
\begin{equation}
\int_{0}^{x^{-}_{2}}dx^{-}\frac{1}{(x^{-}-x^{-}_{3})^2}\left( \frac{(x^{-}_{2}-x^{-})(Z^2-x^{-}x_{2}^{+})}{(x^{-}_{3}-x^{-})^2(Z^2-x_{2}^{-}x_{2}^{+})}\right)^{n-1} \sim \frac{1}{(x_{23}^{-})^n}\left(\frac{x_{2}^{-}}{x_{3}^{-}}\right)^{n}g_{n}(Y)
\label{bulkgny}
\end{equation}
where 
\begin{equation}
Y=\frac{Z^2 x_{23}^{-}}{x^{-}_{3}(x^{+}_{2}x^{-}_{2} - Z^2)}
\end{equation}
We know this is true for $n=1,2$ from \cite{Anand:2017dav, Chen:2019hdv} and we have checked explicitly that it also holds for $n=3,4$.  It must hold for all $n$ since the bulk equations of motion (\ref{eomsec}) are satisfied, but we have
not found a direct proof of this fact.  Curiously the result (\ref{bulk3pt}) matches the exact answer, as was already noted for $n = 1,2$ in \cite{Anand:2017dav, Chen:2019hdv},
which means the geodesic or large-$\Delta$ approximation is actually exact.

\section{Conclusions\label{sect:conclusions}}
In this paper we have taken steps towards understanding the classification of operators needed to reconstruct bulk scalar fields in the presence of conserved currents in the CFT.
The structure is directly relevant for higher-spin currents, but we expect it also plays a role in constructing higher-point correlators involving gauge fields and gravity. We showed that in AdS${}_3$ one can
build a bulk field as in (\ref{AN}), by smearing operators of the form
\begin{equation}
\label{ops}
\partial_{+}^{l+n} J_{(l+h+n,h)}(x^+,x^-)
\end{equation}
where  $J_{(l+h+n,h)}$ are primary fields with spin $n+l$ and twist $\Delta = 2h$.  The coefficients of these operators can be determined by requiring that the bulk field has well-defined correlators.  It seems possible to bypass determining these
coefficients and instead obtain correlators directly from the CFT by making a simple polynomial ansatz for their analytic structure.  Working with
the space of operators (\ref{ops}) means the resulting correlators necessarily have singularities which can be interpreted as generalized Wilson line dressings that run radially to the boundary.

This simple structure of CFT operators does not persist in higher dimensions. Working in a derivative expansion, from \cite{Kabat:2012av} we know the leading (one derivative) CFT operator one needs to correct a bulk scalar field
is the non-primary scalar $j_{\mu}\partial^{\mu}{\cal O}$.  This is indeed the derivative of a primary spin-one current $j_\mu {\cal O}$ of twist $\Delta + d - 2$. At the next level (with three derivatives) one needs the non-primary scalar 
\begin{equation}
{\cal A}=\frac{1}{d}\nabla^{2}j_{\rho} \partial^{\rho} {\cal O} +\frac{d}{2\Delta+2-d} j_{\rho} \partial^{\rho}\nabla^{2} {\cal O}-\partial_{\mu}j_{\rho} \partial^{\mu}\partial^{\rho} {\cal O}.
\end{equation}
However this is {\em not} a descendant of a twist $\Delta+d-2$ spin-two primary.
At leading order in $1/N$ the only twist $\Delta+d-2$ spin-two symmetric traceless primary is
\begin{equation}
J_{\mu \rho}=(j_{\rho} \partial_{\mu} {\cal O}+j_{\mu} \partial_{\rho} {\cal O})-\frac{\Delta}{d}( \partial_{\mu}  j_{\rho}{\cal O}+ \partial_{\rho} j_{\mu}{\cal O})-\frac{2}{d}\eta_{\mu \rho}j_{\nu}\partial^{\nu} {\cal O}
\end{equation}
and 
\begin{equation}
{\cal A} \neq \partial^{\mu} \partial^{\rho} J_{\mu \rho}.
\end{equation}
So additional conformal families are required to build a bulk field in higher-dimensional AdS.  We leave the classification of these operators to future work.

\bigskip
\goodbreak
\centerline{\bf Acknowledgements}
\noindent
DK is supported by U.S.\ National Science Foundation grant PHY-1820734.  GL is supported in part by the Israel Science Foundation under grant 447/17. 

\appendix
\section{Computing 3-point correlators \label{appendix:geq}}
In this appendix we derive the CFT correlators given in (\ref{phi0a}) and (\ref{An}). 

\subsection{$\langle\phi^{(0)}\overline{\cal O}j_{(n,0)}\rangle$}
For $\Delta >n$ we start with
\begin{eqnarray}
\langle{\cal O}(x_{1}^{+},x_{1}^{-}){\cal O}(x_{2}^{+},x_{2}^{-})j_{(n,0)}(x_{3}^{-})\rangle&=&\frac{\gamma_{n}}{(x_{12}^{-})^{\Delta-n}(x_{12}^{+})^{\Delta}(x_{13}^{-})^{n}(x_{23}^{-})^{n}} \\[5pt]
\nonumber
&=&\frac{\gamma_{n}}{(x_{23}^{-})^{n}}\frac{\partial_{x_{2}^{-}}^{\Delta-n-1}\partial_{x_{2}^{+}}^{\Delta-1}\partial_{x_{3}^{-}}^{n-1}}{\Gamma(n)\Gamma(\Delta)\Gamma(\Delta-n)}\frac{1}{x_{12}^{-}x_{12}^{+}x_{13}^{-}}
\end{eqnarray}
Thus the expression we will eventually have to evaluate is some derivatives acting on 
\begin{equation}
\int_{t'^2+y'^2\leq Z^2} dt'dy'\big(\frac{Z^2-y'^2-t'^2}{Z}\big)^{\Delta-2}\frac{1}{(x_{12}^{-}-iy'+t')(x_{12}^{+}+iy'+t')(x_{13}^{-}-iy'+t')}
\end{equation}
To evaluate this integral we change variables to $t'=r \cos \theta$, $y'=r \sin \theta$, $\alpha= e^{i\theta}$ to get
\begin{equation}
\int_{0}^{Z} rdr\big(\frac{Z^2-r^2}{Z}\big)^{\Delta-2}\oint_{|\alpha|=1} \frac{\alpha d \alpha}{i(\alpha x_{12}^{-}+r)(x_{12}^{+}+r\alpha)(\alpha x_{13}^{-}+r)}
\end{equation}
The integration contour encloses the poles at $\alpha=-\frac{r}{x_{12}^{-}}$ and $\alpha=-\frac{r}{x_{13}^{-}}$,\footnote{This is true in the boundary limit $r \rightarrow 0$.  By analytic continuation we take it to be true everywhere.} giving
\begin{equation}
\int_{0}^{Z} rdr\big(\frac{Z^2-r^2}{Z}\big)^{\Delta-2}\frac{2\pi x_{12}^{+}}{(r^2-x_{12}^{+}x_{12}^{-})(r^2-x_{12}^{+}x_{13}^{-})}.
\end{equation}
Using this, the correlator of interest is
\begin{eqnarray}
& &\langle\phi^{(0)}(Z, x_{1}^{+},x_{1}^{-}){\cal O}(x_{2}^{+},x_{2}^{-})j_{(n,0)}(x_{3}^{-})\rangle\\
\nonumber
&=&\frac{\gamma_{n}}{(x_{23}^{-})^{n}}\frac{\partial_{x_{2}^{-}}^{\Delta-n-1}\partial_{x_{2}^{+}}^{\Delta-1}\partial_{x_{3}^{-}}^{n-1}}{\Gamma(n)\Gamma(\Delta)\Gamma(\Delta-n)}\int_{0}^{Z}rdr\big(\frac{Z^2-r^2}{Z}\big)^{\Delta-2}\frac{x_{12}^{+}}{(r^2-x_{12}^{+}x_{12}^{-})(r^2-x_{12}^{+}x_{13}^{-})}
\end{eqnarray}
Doing the derivatives with respect to $x_{12}^{-}$ and $x_{13}^{-}$ we get
\begin{equation}
\frac{\gamma_{n}}{(x_{23}^{-})^{n}}\frac{\partial_{x_{2}^{+}}^{\Delta-1}}{\Gamma(\Delta)}\int_{0}^{Z}rdr\big(\frac{Z^2-r^2}{Z}\big)^{\Delta-2}\frac{(x_{12}^{+})^{\Delta-1}}{(r^2-x_{12}^{+}x_{12}^{-})^{\Delta-n}(r^2-x_{12}^{+}x_{13}^{-})^{n}}
\end{equation}
Changing variables to $s=\frac{r^2}{x_{12}^{+}}$ and taking the derivative with respect to $x_{12}^{+}$ we get
\begin{equation}
\frac{Z^{\Delta} \gamma_{n}}{(2\Delta-2) (x_{23}^{-})^n (Z^2 - x_{12}^{+}x_{12}^{-})^{\Delta-n} (Z^2 - x_{12}^{+}x_{13}^{-})^{n}}
\end{equation}
Defining $\tilde{\gamma}_n =  {\gamma_n \over 2 \Delta - 2}$ gives (\ref{phi0a}).  For $\Delta \leq n$ one can make a similar computation resulting in the same answer.

\subsection{$\langle{\cal A}^{\Delta,n}_{l} \overline{\cal O} j_{(n,0)}\rangle$}
We start with
\begin{eqnarray}
\langle J_{(l+\frac{\Delta}{2}+n,\frac{\Delta}{2})}(x_{1}^{+},x_{1}^{-}){\cal O}(x_{2}^{+},x_{2}^{-})j_{(n,0)}(x_{3}^{-})\rangle&=&\frac{\alpha^{(n)}_l (x_{23}^{-})^{l}}{(x_{12}^{-})^{\Delta+l}(x_{12}^{+})^{\Delta}(x_{13}^{-})^{l+2n}}\\[5pt]
\nonumber
&=& \alpha^{(n)}_l (x_{23}^{-})^{l}\frac{\partial_{x_{2}^{-}}^{\Delta+l-1}\partial_{x_{2}^{+}}^{\Delta-1}\partial_{x_{3}^{-}}^{l+2n-1}}{\Gamma(l+2n)\Gamma(\Delta)\Gamma(l+\Delta)}\frac{1}{x_{12}^{-}x_{12}^{+}x_{13}^{-}} 
\end{eqnarray}
This gives the 3-point function
\begin{eqnarray}
&&\langle{\cal A}^{(\Delta,n)}_{l}(Z, x_{1}^{+},x_{1}^{-}){\cal O}(x_{2}^{+},x_{2}^{-})j_{(n,0)}(x_{3}^{-})\rangle \\[5pt]
\nonumber
&&=\alpha^{(n)}_l(x_{23}^{-})^{l}\frac{\partial_{x_{2}^{-}}^{l+\Delta-1}\partial_{x_{2}^{+}}^{l+n+\Delta-1}\partial_{x_{3}^{-}}^{l+2n-1}}{\Gamma(l+2n)\Gamma(\Delta)\Gamma(l+\Delta)}\int_{0}^{Z}rdr \left(\frac{Z^2-r^2}{Z}\right)^{2l+2n+\Delta-2}\frac{x_{12}^{+}}{(r^2-x_{12}^{+}x_{12}^{-})(r^2-x_{12}^{+}x_{13}^{-})}
\end{eqnarray}
We first take the derivatives with respect to $x_{2}^{+}$ and $x_{3}^{-}$, then change the integration variable from $r$ to $s=\frac{r^2}{x_{12}^{+}}$ to obtain
\begin{equation}
\frac{(-1)^{\Delta}\alpha^{(n)}_l(x_{23}^{-})^{l}}{2\Gamma(\Delta)}\partial_{x_{2}^{+}}^{+n+\Delta-1}\int_{0}^{\frac{Z^2}{x_{12}^{+}}} ds \left(\frac{Z^2-x_{12}^{+}s}{Z}\right)^{2l+2n+\Delta-2}\frac{1}{(s-x_{12}^{-})^{l+\Delta}(s-x_{13}^{-})^{l+2n}}
\end{equation}
Next we take the derivative with respect to $x_{2}^{+}$ and change the integration variable to $t=\frac{x_{12}^{+} s}{Z^2}$, which gives
\begin{equation}
\frac{\alpha^{(n)}_l\Gamma(2l+2n+\Delta-1)(\chi_{2}-\chi_{1})^{l}(x_{12}^{+})^{n}}{2\Gamma(\Delta)\Gamma(l+n)Z^{\Delta+2n}\chi_{1}^{l+\Delta}\chi_{2}^{l+2n}}\int_{0}^{1}dt \frac{t^{l+n+\Delta-1}(1-t)^{l+n-1}}{\big(1-\frac{t}{\chi_{1}}\big)^{l+\Delta}\big(1-\frac{t}{\chi_{2}}\big)^{l+2n}}
\end{equation}
where
\begin{equation}
\chi_{1}=\frac{x_{12}^{+}x_{12}^{-}}{Z^2}\,, \qquad \chi_{2}=\frac{x_{12}^{+}x_{13}^{-}}{Z^2}
\end{equation}
The integral over $t$ is
\begin{equation}
B(n+l,l+n+\Delta)(1-\frac{1}{\chi_{2}})^{-(l+n+\Delta)}F(l+n+\Delta,l+\Delta,2l+2n+\Delta,\frac{\chi_{2}-\chi_{1}}{\chi_{1}(\chi_{2}-1)})
\end{equation}
Defining $\tilde{\alpha}^{(n)}_{l}=\frac{\alpha^{(n)}_l\Gamma(2l+2n+\Delta-1)B(n+l,l+n+\Delta)}{2\Gamma(l+n)\Gamma(\Delta)}$ and ${\cal X}=\frac{\chi_{2}-\chi_{1}}{\chi_{1}(\chi_{2}-1)}$, the 3-point function can be written as
\begin{equation}
\tilde{\alpha}^{(n)}_{l}\frac{(x_{12}^{+})^{n}\chi_{2}^{\Delta-n}}{Z^{\Delta+2n}\chi_{1}^{\Delta}(\chi_{2}-1)^{\Delta+n}}{\cal X}^{l}F(l+n+\Delta,l+\Delta,2l+2n+\Delta,{\cal X})
\end{equation}
Using the hypergeometric identity
\begin{equation}
F(a,b,c,z)=(1-z)^{-b}F(b,c-a,c,\frac{z}{z-1}),
\end{equation}
and defining $Y=\frac{{\cal X}}{{\cal X}-1}$, this can be written as equation (\ref{An}).

\section{Higher spin primaries \label{appendix:hsp}}
Here we show how to construct higher spin primary\footnote{under global conformal transformations} operators from a scalar and a conserved current.
We start with CFT${}_2$. It is easy to build operators that transform correctly under dilatations and Lorentz transformations, the problem is with special conformal transformations.
A primary $J_{(l+\frac{\Delta}{2}+n,\frac{\Delta}{2})}$ (with $l\geq 0$) of conformal weights $(l+\frac{\Delta}{2}+n,\frac{\Delta}{2})$ transforms under $L_{1}$ and
$\overline{L}_{1}$ as
\begin{eqnarray}
L_{1} J_{(l+\frac{\Delta}{2}+n,\frac{\Delta}{2})}(x^{+},x^{-})&=&2(l+\frac{\Delta}{2}+n)x^{-}J_{(l+\frac{\Delta}{2}+n,\frac{\Delta}{2})} +(x^{-})^2\partial_{-}J_{(l+\frac{\Delta}{2}+n,\frac{\Delta}{2})}\nonumber \\
\overline{L}_{1} J_{(l+\frac{\Delta}{2}+n,\frac{\Delta}{2})}(x^{+},x^{-})&=&2({\Delta \over 2})x^{+} J_{(l+\frac{\Delta}{2}+n,\frac{\Delta}{2})} +(x^{+})^2\partial_{+}J_{(l+\frac{\Delta}{2}+n,\frac{\Delta}{2})}.
\end{eqnarray}
Using the transformation (where ${\cal O}$ has dimension $(h,h)$, with $h = \Delta / 2$)
\begin{eqnarray}
&&L_{1} {\cal O}=2hx^{-}{\cal O}+(x^{-})^2 \partial_{-} {\cal O} \nonumber \\
&&\overline{L}_{1} {\cal O}=2h x^{+} {\cal O}+(x^{+})^2 \partial_{+} {\cal O} \nonumber \\
&&L_{1} j_{(n,0)}=2nx^{-} j_{(n,0)}+(x^{-})^2 \partial_{-} j_{(n,0)} \nonumber \\
&&\overline{L}_{1} j_{(n,0)} = (x^{+})^2 \partial_{+} j_{(n,0)}
\end{eqnarray}
it is simple to show that one can write (to leading order in $1/N$) a spin $l+n$ primary operator with conformal dimensions $(l+\frac{\Delta}{2}+n,\frac{\Delta}{2})$ as 
\begin{eqnarray}
&&J_{(l+\frac{\Delta}{2}+n,\frac{\Delta}{2})}=\sum_{k=0}^{l} d_{k} \partial_{-}^{k} j_{-} \partial^{l-k}_{-} {\cal O}\nonumber\\
&&d_{k} = \frac{(-1)^{k}}{\Gamma(k+1)\Gamma(l-k+1)\Gamma(k+2n)\Gamma(l-k+\Delta)}
\end{eqnarray}

To see that these operators play a role in bulk reconstruction, note that they have scalar descendants $\partial^{l+n}_{+}J_{(l+\frac{\Delta}{2}+n,\frac{\Delta}{2})}$ which
behave under special conformal transformations as
\begin{eqnarray}
{\overline L}_{1}(\partial^{l+n}_{+}J_{(l+\frac{\Delta}{2}+n,\frac{\Delta}{2})}) &\sim& (\hbox{\rm primary result}) + \partial^{l+n-1}_{+}J_{(l+\frac{\Delta}{2}+n,\frac{\Delta}{2})} \nonumber \\
L_1 (\partial^{l+n}_{+} J_{(l+\frac{\Delta}{2}+n,\frac{\Delta}{2})}) &\sim& (\hbox{\rm primary result})
\end{eqnarray}
where (primary result) is the appropriate transformation for a primary scalar of dimension $(l+\frac{\Delta}{2}+n,l+\frac{\Delta}{2}+n)$.
 
This matches expectations for building corrections to a bulk scalar field interacting with a gauge field \cite{Kabat:2013wga}. In that paper it was shown that
in holographic gauge under a special conformal transformation with parameter $b_{\mu}$ the non-primary scalars needed to correct the bulk scalar field transform as
\begin{equation}
\delta_{b} {\cal O} \sim (\hbox{\rm primary scalar result}) + b^{\mu} \partial \cdots \partial j_{\mu} \partial \cdots \partial {\cal O}.
\end{equation}
In 2-d CFT with only $j_{-}$ present this means the transformation of ${\cal O}$ under $L_{1}$ is that of a scalar, but under $\overline{L}_{1}$ is not.

A similar construction is available in higher dimensions. For instance given a primary scalar of dimension $\Delta$ and a conserved current $j_\mu$ in a d-dimensional CFT the combination
\begin{equation}
J_{\rho \mu}=(j_{\rho}\partial_{\mu} {\cal O}+j_{\mu}\partial_{\rho} {\cal O})-\frac{\Delta}{d}(\partial_{\mu}j_{\rho} {\cal O}+\partial_{\rho}j_{\mu} {\cal O})-\frac{2}{d}(\eta_{\mu \rho}j_{\alpha}\partial^{\alpha} {\cal O})
\end{equation}
is a primary spin-2 symmetric traceless operator at leading order in $1/N$.

\section{Determining operator coefficients for $n=1$ \label{appendix:n=1}}
The double-trace operators should not introduce any additional branch cuts and should cancel the ill-defined pole which is present in the zeroth-order correlator.
In this section we show -- in the simplest setting of $n = 1$ -- how this requirement can be used to determine the coefficients of the double-trace operators.
 
We start with the sum (\ref{3pointphi1}) of the correlators of the double-trace operators ${\cal A}_l^{\Delta,n}$ weighted by coefficients $a_{l}$.
\begin{equation}
\sum_{l=0}^{\infty} \frac{a_{l}\tilde{\alpha}^{(n)}_{l}}{(x_{23}^{-})^{n}Z^{\Delta}(1 - \chi_1)^{\Delta}}\left(\frac{\chi_1}{\chi_2}\right)^{n}\left(\frac{Y}{Y-1}\right)^{n}(-Y)^{l}F(l+\Delta,l+n, 2l+2n+\Delta,Y)
\label{suman}
\end{equation}
The hypergeometric function has a branch cut for $Y>1$ whose $i\epsilon$ prescription is ill-defined due to its dependence on $1-\chi_{2}$.
So we need to make sure that after summing the tower of double-trace operators the branch cuts cancel. To do this we look at the discontinuity across the cut of (\ref{suman}).
Using
\begin{eqnarray}
&&F(a,b,c,z+i\epsilon)- F(a,b,c,z-i\epsilon)=\frac{2\pi i \Gamma(c)(z-1)^{c-a-b}}{\Gamma(a)\Gamma(b)\Gamma(1+c-a-b)}F(c-a,c-b,1+c-a-b,1-z) \nonumber \\
&&F(a,b,c,z)=(1-z)^{-b}F\big(b,c-a,c,\frac{z}{z-1}\big)
\end{eqnarray}
the requirement that the sum of the discontinuities across the cut vanishes becomes the requirement that
\begin{equation}
\sum_{l=0}^{\infty} a_{l}\tilde{\alpha}^{(n)}_{l} (-1)^{l} \frac{\Gamma(2l+\Delta+2n)}{\Gamma(l+\Delta)\Gamma(l+n)}F(l+\Delta+n,1-l-n,1+n,z)=0.
\end{equation}

As an example, we solve this for the case $n=1$. We are looking for coefficients $a_{l}\tilde{\alpha}^{(n=1)}_{l}$ such that 
\begin{equation}
\sum_{l=0}^{\infty} a_{l}\tilde{\alpha}^{(n=1)}_{l}(-1)^{l} \frac{\Gamma(2l+\Delta+2)}{\Gamma(l+\Delta)\Gamma(l+1)}F(l+\Delta+1,-l,2,z)=0.
\label{coefn1}
\end{equation}
We start with the identity (taken from equation (277) of \cite{Kabat:2015swa})
\begin{equation}
F(\Delta,1,2,z)=\sum_{l=0}^{\infty} (-1)^{l}\frac{\Gamma(l+\Delta)(2l+\Delta+1)}{\Gamma(\Delta)\Gamma(l+2)}F(l+\Delta +1,-l,2,z)
\label{suman1}
\end{equation}
and act on both sides with the differential operator
\begin{equation}
{\cal L}_{\Delta}=z(1-z)\frac{d^2}{d^2 z}+(2-(\Delta+2)z)\frac{d}{dz}-\Delta
\end{equation}
The hypergeometric functions appearing in (\ref{suman1}) are all eigenfunctions of this operator and we get
\begin{equation}
0=\sum_{l=0}^{\infty} (-1)^{l} \frac{\Gamma(l+\Delta+1)(2l+\Delta+1)}{\Gamma(\Delta)\Gamma(l+1)}F(l+\Delta +1,-l,2,z).
\end{equation}
Comparing to (\ref{coefn1}) we see that
\begin{equation}
\label{aalpha}
a_{l}\tilde{\alpha}^{(n=1)}_{l}=c_{1}\frac{\Gamma(l+\Delta)\Gamma(l+\Delta+1)}{\Gamma(2l+\Delta+1)}. 
\end{equation}
for some constant $c_1$.  It is important to remember that these coefficients are not unique. For instance acting again with ${\cal L}_{\Delta}$ will give a different
set of coefficients.  All of these choices will cancel the branch cut (at least formally) but won't necessarily suffice to cancel the ill-defined pole in the $\phi^{(0)}$ correlator.

To proceed we use these coefficients to sum the contribution of the double-trace operators.  This will give us an expression for the correction to the correlator (\ref{3pointphi1})
for a particular choice of $f_1(Y)$.  We know it won't have branch cuts, but we still have to check whether it will cancel the ill-defined pole in the zeroth-order correlator (\ref{phi0}).
With the coefficients (\ref{aalpha}) the expression for $f_{n=1}(Y)$ is
\begin{equation}
f_{n=1}(Y)=c_1\sum_{l=0}^{\infty}\frac{\Gamma(l+\Delta)\Gamma(l+\Delta+1)}{\Gamma(2l+\Delta+1)}(-Y)^{l}F(l+\Delta,l+1,2l+\Delta+2,Y)
\end{equation}
We can evaluate this using \cite{hansen}\footnote{This formula is true for $x<1$ which is fine for us and $a\leq \frac{1}{2}$ which we will ignore.}
\begin{equation}
\sum_{k=0}^{\infty} \frac{(a)_{k} (b)_{k} (c)_{k}}{\Gamma(k+1)(c)_{2k}} x^{k} F(a+k,b+k,c+2k+1,-x)=1.
\end{equation}
This gives
\begin{equation}
f_{n=1}(Y) = c_1 \Gamma(\Delta)
\end{equation}
So if we choose $c_1=\tilde{\gamma}_{1}/\Gamma(\Delta)$ then we recover (\ref{3pointphi}) with $g_{n=1}(Y)=1$.  For $n = 1$ this was the minimal choice for
$g_n(Y)$ that made the corrected bulk correlator well-defined.
 
\section{$\phi^{(1)}$ from bulk equations of motion \label{appendix:bulkeom}}
In the section we show how to obtain the bulk operator by solving bulk equations of motion. We do this for a massless bulk scalar coupled to a Chern-Simons
gauge field.  Thus in the CFT we have a $\Delta=2$ primary scalar ${\cal O}$ and chiral spin-one primary current $j_{-}(x^{-})$.  Our starting point is the zeroth-order
3-point function (\ref{phi0Oj}).

To cancel the ambiguous singularities in the correlator (\ref{phi0Oj}) we have to add a tower of higher-dimension double-trace operators to our definition of the bulk scalar.
The available ingredients are the operators
\[
{\cal O}_{nm} = \partial_-^n j_-(x^-) \, \partial_-^m \partial_+^{n+m+1} {\cal O}(x^+,x^-) \qquad n,m = 0,1,2,\ldots
\]
These operators are Lorentz scalars on the boundary, and they have scaling dimension $\Delta_{nm} = 2n + 2m + 4$, but we have not imposed any kind of conformal primary condition.
We will write the correction to the bulk field as\footnote{In this appendix $K_{nm}$ is normalized slightly differently from $K_\Delta$, $K_{nm} = (2\Delta_{nm} - 2) K_{\Delta_{nm}}$.  Sorry about that.}
\begin{eqnarray}
\label{HDOsum}
&& \phi^{(1)}(t,x,Z) = \sum_{n,m=0}^\infty a_{nm} \int K_{nm} {\cal O}_{nm} \\
\nonumber
&& = \sum_{n,m=0}^\infty a_{nm} {2n + 2m + 3 \over \pi} \hspace{-3mm} \int\limits_{\hspace{5mm} (t')^2 + (y')^2 < Z^2} \hspace{-5mm} dt'dy' \left({Z^2 - (t')^2 - (y')^2 \over Z}\right)^{2n+2m+2}\, {\cal O}_{nm}(t + t',x + iy')
\end{eqnarray}
where we've written the smearing integral explicitly in the second line.

To determine the coefficients $a_{nm}$ we will use the bulk equations of motion.\footnote{This approach was originally used in \cite{Kabat:2011rz, Heemskerk:2012mn}.}  For a massless scalar of charge $q \sim 1/N$ these read
\begin{equation}
{1 \over \sqrt{-g}} D_M \sqrt{-g} g^{MN} D_N \phi = 0
\end{equation}
where $D_M = \partial_M + iqA_M$.\footnote{Our notation is that $D_M$ is a gauge covariant but not AdS covariant derivative.  That is, it has the gauge connection but not the Christoffel connection.
Below $\nabla_M$ will denote a derivative that is AdS covariant but not gauge covariant.}
Expanding in powers of $q$ we write the fields as $\phi = \phi^{(0)} + \phi^{(1)} + \cdots$, $A_M = A_M^{(0)} + A_M^{(1)} + \cdots$.  At lowest order the scalar field obeys
\begin{equation}
{1 \over \sqrt{-g}} \partial_M \sqrt{-g} g^{MN} \partial_N \phi^{(0)} = 0
\end{equation}
which is already satisfied by (\ref{masslessphi0}).  At lowest order the gauge field is given by
\begin{equation}
\label{Anot}
A^{(0)}_+ = 0 \qquad A^{(0)}_- = j_-(x^-) \qquad A^{(0)}_Z = 0
\end{equation}
That is, the light-cone components of the bulk gauge field are equal to the current on the boundary -- a property special to Chern-Simons -- while the
$Z$ component is set to zero by our gauge choice.  It follows that the divergence of $A^{(0)}$ vanishes,
\begin{equation}
\nabla_M A^{M\,(0)} = {1 \over \sqrt{-g}} \partial_M \sqrt{-g} A^{M\,(0)} = 0
\end{equation}
which simplifies the scalar equation of motion.  At first order we just have to solve
\begin{equation}
\label{bulkeom}
{1 \over \sqrt{-g}} \partial_M \sqrt{-g} g^{MN} \partial_N \phi^{(1)} = 4 i q {Z^2 \over R^2} A_-^{(0)} \partial_+ \phi^{(0)}
\end{equation}

First let's evaluate the left hand side of (\ref{bulkeom}).  We plug in our ansatz (\ref{HDOsum}) and use the fact that the free fields
$\phi^{(0)}_{nm} = \int K_{nm} {\cal O}_{nm}$ are eigenfunctions of the Laplacian,
\begin{equation}
{1 \over \sqrt{-g}} \partial_M \sqrt{-g} g^{MN} \partial_N \phi^{(0)}_{nm} = m_{nm}^2 \phi^{(0)}_{nm}
\end{equation}
where the $({\rm mass})^2$ is $m_{nm}^2 = \Delta_{nm} (\Delta_{nm} - 2)$.  Thus we have
\begin{eqnarray}
&& {1 \over \sqrt{-g}} \partial_M \sqrt{-g} g^{MN} \partial_N \phi^{(1)} = \sum_{n,m=0}^\infty a_{nm} {1 \over \pi}
(2n+2m+2)(2n+2m+3)(2n+2m+4) \\
\nonumber
&& \int\limits_{\hspace{5mm} (t')^2 + (y')^2 < Z^2} \hspace{-5mm} dt'dy' \left({Z^2 - (t')^2 - (y')^2 \over Z}\right)^{2n+2m+2}\,
\partial_-^n j_-(x^- + \overline{w}) \, \partial_-^m \partial_+^{n+m+1} {\cal O}(x^++w,x^-+\overline{w})
\end{eqnarray}
where $w = t' + iy'$, $\overline{w} = t' - iy'$.  Next we Taylor expand the operators about the point $(x^+,x^-)$ and do the smearing integral.
This ends up producing an expansion of the left hand side in powers of $z$.  Introducing new variables for the sums $N$, $M$ that correspond to fixed numbers of derivative
operators we find
\begin{eqnarray}
\label{lhs}
&& \nabla_M \nabla^M \phi^{(1)} = \sum_{N = 0}^\infty \sum_{M = 0}^\infty c_{NM} z^{2N+2M+4} \partial_-^N j_-(x^-) \, \partial_-^M \partial_+^{N+M+1} {\cal O}(x^+,x^-) \\
\label{lhs2}
&& c_{NM} = \sum_{n = 0}^N \sum_{m = 0}^M a_{nm} {2 \, (n+m+1) \, \Gamma(2n + 2m + 5) \over (N-n)! \, (M-m)! \, \Gamma(N+M+n+m+4)}
\end{eqnarray}

On the right hand side of (\ref{bulkeom}) we perform an analogous expansion.  Plugging (\ref{masslessphi0}) and (\ref{Anot}) into the right hand side of
(\ref{bulkeom}) gives
\begin{equation}
4 i q {Z^2 \over R^2}  j_-(x^-) \partial_+ {1 \over \pi}  \hspace{-3mm} \int\limits_{\hspace{5mm} (t')^2 + (y')^2 < Z^2} \hspace{-5mm} dt'dy' \, {\cal O}(x^++w,x^-+\overline{w})
\end{equation}
Again we Taylor expand the operator about the point $(x^+,x^-)$ and do the smearing integral.  This gives
\begin{equation}
\label{rhs}
{4 i q \over R^2} \sum_{M=0}^\infty Z^{2M+4} {1 \over M! \, (M+1)!} j_-(x^-) \partial_-^M \partial_+^{M+1} {\cal O}(x^+,x^-)
\end{equation}

Comparing (\ref{lhs}) and (\ref{rhs}), with $c_{NM}$ defined in (\ref{lhs2}), the bulk equations of motion give us the following system of equations to determine $a_{nm}$.
\begin{equation}
c_{NM} = \left\lbrace\begin{array}{ll}
{\displaystyle 4 i q \over \displaystyle R^2} {\displaystyle 1 \over \displaystyle M! \, (M+1)!} & \quad \hbox{\rm for $N = 0$} \\
\,\,\, 0 &  \quad \hbox{\rm for N = 1,2,3,\ldots}
\end{array}\right.
\end{equation}
After a bit of guesswork the solution to this system is
\begin{equation}
\label{anm}
a_{nm} = {i q \over R^2} \, {(-1)^n (n+m)! \, (n + m + 1)! \over n! \, m! \, (n+1)! \, (m+1)! (2n + 2m + 2)!}
\end{equation}

\subsection{Computing $\langle \phi^{(1)} {\cal O} j_- \rangle$}
Now let's calculate the correlator $\langle \phi^{(1)} {\cal O} j_- \rangle$ where the expression for $\phi^{(1)}$ as a sum of double-trace operators is given in (\ref{HDOsum})
and the coefficients $a_{nm}$ are given in (\ref{anm}).  Unlike the body of the paper, where we determined correlators using analyticity, here we will obtain the corrected
correlator by directly summing the contribution of the double-trace operators.

First note that by large-$N$ factorization
\begin{eqnarray}
\langle \big({\cal O}(x_1) j_-(x_1)\big) {\cal O}(x_2) j_-(x_3) \rangle & = & \langle {\cal O}(x_1) {\cal O}(x_2) \rangle \langle j_-(x_1) j_-(x_3) \rangle \\
\nonumber
& = & - {1 \over (x_{12}^+)^2 (x_{12}^-)^2 (x_{13}^-)^2}
\end{eqnarray}
Using this the correlator of $\phi^{(1)}$ with ${\cal O}$ and $j_-$ is
\begin{eqnarray}
\langle \phi^{(1)}(x_1,Z) {\cal O}(x_2) j_-(x_3) \rangle & = & \sum_{n,m=0}^\infty a_{nm} {2n + 2m + 3 \over \pi} \partial_{x_2^+}^{n+m+2} \partial_{x_2^-}^{m+1} \partial_{x_3^-}^{n+1} \\
\nonumber
&&\hspace{-15mm} \int\limits_{\hspace{5mm} (t')^2 + (y')^2 < Z^2} \hspace{-5mm} dt'dy' \left({Z^2 - (t')^2 - (y')^2 \over Z}\right)^{2n+2m+2}\, {1 \over (x_{12}^+ + w)(x_{12}^- + \overline{w})(x_{13}^- + \overline{w})}
\end{eqnarray}
where $w = t + t' + i y'$, $\overline{w} = t' - i y'$ and we've pulled out some derivatives so the integrand has simple poles.  We set $t' = r \cos \theta$, $y' = r \sin \theta$ and let $\alpha = e^{i \theta}$ so that
$w = r \alpha$, $\overline{w} = r / \alpha$ and the integration measure becomes
\begin{equation}
\int\limits_{\hspace{5mm} (t')^2 + (y')^2 < Z^2} \hspace{-5mm} dt'dy' = \int_0^Z rdr \oint\limits_{\vert \alpha = 1} {d\alpha \over i \alpha}
\end{equation}
The contour integral over $\alpha$ leads to
\begin{eqnarray}
\langle \phi^{(1)} {\cal O} j_- \rangle & = & \sum_{n,m=0}^\infty a_{nm} {2n + 2m + 3 \over \pi} \partial_{x_2^+}^{n+m+2} \partial_{x_2^-}^{m+1} \partial_{x_3^-}^{n+1} \\
\nonumber
& & 2 \pi x_{12}^+ \int_0^Z rdr \, \left({z^2 - r^2 \over z}\right)^{2n+2m+2} {1 \over (r^2 - x_{12}^+ x_{12}^-)(r^2 - x_{12}^+ x_{13}^-)}
\end{eqnarray}
Next we act with the $x_2^-$ and $x_3^-$ derivatives, and after that we rearrange the sums by introducing $N = n + m$.  That is, we rearrange the sums using
\begin{equation}
\sum_{n,m=0}^\infty S(n,m) = \sum_{N = 0}^\infty \sum_{m = 0}^N S(N-m,m)
\end{equation}
The sum over $m$ is a simple binomial and leads to the following expression for $\langle\phi^{(1)} {\cal O} j_-\rangle$.
\begin{equation}
{i q \over R^2} \sum_{N=0}^\infty {(N+1)! \, (2N+3) \over (2N+2)!} \partial_{x_2^+}^{N+2} \int_0^Z 2 r dr  \, \left({Z^2 - r^2 \over Z}\right)^{2N+2} {(x_{12}^+)^{2N+3} (x_{23}^-)^N \over (r^2 - x_{12}^+ x_{12}^-)^{N+2} (r^2 - x_{12}^+ x_{13}^-)^{N+2}}
\end{equation}
Next it's convenient to set $r^2 = x_{12}^+s$ to obtain
\begin{equation}
{i q \over R^2} \sum_{N=0}^\infty {(N+1)! \, (2N+3) \over (2N+2)!} \partial_{x_2^+}^{N+2} \int_0^{Z^2/x_{12}^+} ds  \, \left({Z^2 - x_{12}^+ s \over Z}\right)^{2N+2} {(x_{23}^-)^N \over (s - x_{12}^-)^{N+2} (s - x_{13}^-)^{N+2}}
\end{equation}
Then we act with the $x_2^+$ derivatives\footnote{There's no contribution from the $x_2^+$ derivative acting on the upper limit of integration since the integrand vanishes there.} and rescale $s = Z^2 t / x_{12}^+$ to obtain
\begin{equation}
{i q \over R^2} {x_{12}^+ \over z^4} \sum_{N=0}^\infty (N+1) (2N+3) \int_0^1 dt \, t^{N+2} (1-t)^N {(\chi_2 - \chi_1)^N \over (t-\chi_1)^{N+2} (t - \chi_2)^{N+2}}
\end{equation}
where we've introduced the translation-, Lorentz- and scale-invariant combinations\footnote{A little parameter counting: the correlator $\langle \phi^{(1)} O j_- \rangle$ depends on one bulk point and two boundary points,
so naively has 7 parameters.  The current $j_-(x^-)$ is chiral which reduces the count to 6.  Translation invariance in the CFT reduces it to the 4 combinations $x_{12}^+$, $x_{12}^-$, $x_{13}^-$, $Z$, Lorentz invariance further restricts to
the 3 combinations $x_{12}^+ x_{12}^-$, $x_{12}^+ x_{13}^-$, $Z$ and finally imposing scale invariance leaves the two combinations $\chi_1$, $\chi_2$.  Note that we haven't imposed invariance under special conformal transformations.}
\begin{equation}
\chi_1 = {x_{12}^+ x_{12}^- \over Z^2} \qquad\qquad \chi_2 = {x_{12}^+ x_{13}^- \over Z^2}
\end{equation}
Summing over $N$ and integrating over $t$ we are left with the surprisingly simple result
\begin{equation}
\langle \phi^{(1)}(x_1,Z) {\cal O}(x_2) j_-(x_3) \rangle = {i q \over R^2} {x_{12}^+ \over Z^4} {1 \over \big(\chi_1 - 1\big)^2 (\chi_2-1) \chi_2}
\end{equation}

Let's combine this with our lowest-order result (\ref{phi0Oj}), which can be written in the form
\begin{equation}
\langle \phi^{(0)}(x_1,Z) {\cal O}(x_2) j_-(x_3) \rangle = {\lambda \over 2} {x_{12}^+ \over Z^4} {1 \over (\chi_1 - 1)(\chi_2-1) (\chi_1 - \chi_2)}
\end{equation}
With the normalization  ${\lambda \over 2} = - {i q \over R^2}$ we get
\begin{eqnarray}
\nonumber
\langle \left(\phi^{(0)} + \phi^{(1)}\right) {\cal O} j_- \rangle & = & - {i q \over R^2} {x_{12}^+ \over Z^4} {\chi_1 \over (\chi_1 - 1)^2 (\chi_1-\chi_2) \chi_2} \\
& = & {i q \over R^2} {Z^2 x_{12}^- \over x_{13}^- x_{23}^-} {1 \over \big(x_{12}^+ x_{12}^- - Z^2\big)^2}
\end{eqnarray}
This final result has the expected form.  It agrees with (\ref{rewrite3pointphi}) for $n = 1$ with $g_1(Y) = 1$.  There's a singularity at $x_{12}^+ x_{12}^- = Z^2$, where the bulk operator $\phi = \phi^{(0)} + \phi^{(1)}$ is null separated from the boundary operator ${\cal O}$.
There are also poles when $x_1^- = x_3^-$ and $x_2^- = x_3^-$.  These poles give $\delta$-function commutators with the current and show that charge is located on the boundary at $x^- = x_1^-$
and $x^- = x_2^-$.

\section{Building general polynomial $f_{n}(Y)$ \label{appendix:coef}}
We wish to show that we can choose coefficients $a_{l}$ so that
\begin{equation}
f_n(Y) = \sum_{l=0}^{\infty} a_{l} (-Y)^{l}F(l+\Delta, l+n, 2l+\Delta+2n,Y)
\end{equation}
is equal to any polynomial we desire.
A useful starting point is the identity
\begin{equation}
\sum_{k=0}^{\infty} \frac{(\alpha)_{k} (\beta)_{k} (\gamma)_{k}}{\Gamma(k+1)(\gamma)_{2k}} x^{k} F(\alpha+k,\beta+k,\gamma+2k+1,-x)=1
\end{equation}
This formula is true for $x<1$ which is fine for us and $a\leq \frac{1}{2}$ which we will ignore.
Let us label 
\begin{equation}
d^{(\alpha,\beta,\gamma)}_{k}= \frac{(\alpha)_{k} (\beta)_{k} (\gamma)_{k}}{\Gamma(k+1)(\gamma)_{2k}}
\end{equation}
Clearly getting $f_{n}(Y)=1$ is possible simply by picking
\begin{equation}
a_{l}=d_{l}^{(\Delta,n,2n+\Delta-1)} \qquad {\rm for} \,\, l \geq 0.
\end{equation}
Now suppose we are looking for coefficients $a_{l}$ such that 
\begin{equation}
\sum_{l=0}^{\infty} a_{l} (-Y)^{l}F(l+\Delta, l+n, 2l+\Delta+2n,Y)=Y
\end{equation}
If we take $a_{0}=0$ then this becomes
\begin{equation}
\sum_{l=1}^{\infty} a_{l} (-Y)^{l-1}F(l+\Delta, l+n, 2l+\Delta+2n,Y)=-1
\end{equation}
which can be satisfied by setting
\begin{equation}
a_{l}=-d_{l-1}^{(\Delta+1,n+1,2n+\Delta+1)} \qquad {\rm for} \,\, l \geq 1.
\end{equation}
Clearly this procedure can be repeated to obtain any positive power of $Y$ we want.

\section{Minimal solution for $g_n(Y)$ \label{appendix:gnY}}
Here we solve the system of equations (\ref{gpolycon}) to find the minimal polynomial $g_n(Y)$ and we show that the resulting correlator satisfies the bulk equations of motion (\ref{eomsec}).

\subsection{Obtaining $g_n(Y)$}
We wish to solve
\begin{equation}
\sum_{j=0}^{m}(-1)^{j}c_{m-j}{n \choose j}=0
\end{equation}
with $c_0=1$.
To find a solution we use the Chu-Vandermonde identity
\begin{equation}
\sum_{k=0}^{r} {q \choose k}{p \choose r-k}={q+p \choose r}
\end{equation}
together with ${0 \choose r}=\delta_{0r}$.
We see that by setting
\begin{equation}
c_{m-j}=(-1)^{j}{-n \choose m-j}
\end{equation}
the equation will be obeyed. This can be rewritten using the negated upper index binomial coefficient identity 
\begin{equation}
{n \choose k}=(-1)^{k}{k-n-1 \choose k}
\end{equation}
to obtain
\begin{equation}
c_{m-j}={n+m-j-1 \choose m-j}
\end{equation}

\subsection{Checking bulk equations of motion\label{appendix:eom}}
To check the equation of motion (\ref{eomsec}) we first compute (remember $l=0 \ldots n-1$)
\begin{equation}
\big(\nabla^2_{x_1,Z}-\Delta(\Delta-2)\big)\left((x_{23}^{-})^{-n} \frac{Z^{\Delta}}{(Z^2-x_{12}^{-}x_{12}^{+})^{\Delta}}\Big(\frac{\chi_{1}}{\chi_{2}}\Big)^{n}Y^{l}\right)
\label{eomapp}
\end{equation}
After some algebra this is given by
\begin{equation}
4l(\Delta+l-1)\frac{Z^{\Delta+2l}}{(Z^2-x_{12}^{-}x_{12}^{+})^{\Delta+l}}\frac{(x_{23}^{-})^{l-n}(x_{12}^{-})^{n}}{(x_{13}^{-})^{l+n}} - 4(n+l)(\Delta+l)\frac{Z^{\Delta+2l+2}}{(Z^2-x_{12}^{-}x_{12}^{+})^{\Delta+l+1}}\frac{(x_{23}^{-})^{l+1-n}(x_{12}^{-})^{n}}{(x_{13}^{-})^{l+n+1}}
\label{eomapp1}
\end{equation}
Note the the two terms are related up to a coefficient by $l \rightarrow l+1$.
The right-hand side of the equation of motion (\ref{eomsec}) is
\begin{equation}
\sim \frac{Z^{\Delta+2n}}{(Z^2-x_{12}^{-}x_{12}^{+})^{\Delta+n}}\frac{(x_{12}^{-})^{n}}{(x_{13}^{-})^{2n}}
\end{equation}
We see that this is just the second term in (\ref{eomapp1}) for $l=n-1$. So all other terms must cancel. Since $g_{n}(Y)=\sum_{l=0}^{n-1} c_{l}Y^{l}$ this means that the $c_{l}$ should obey
\begin{equation}
c_{l+1}=\frac{n+l}{l+1}c_{l}
\end{equation}
which is indeed obeyed by $c_{l}={n+l-1 \choose l}$.

\providecommand{\href}[2]{#2}\begingroup\raggedright\endgroup

\end{document}